\documentclass[conference]{IEEEtran}
\IEEEoverridecommandlockouts
\usepackage{cite}
\usepackage{amsmath,amssymb,amsfonts}
\usepackage{algorithmic}
\usepackage{graphicx}
\usepackage{textcomp}
\usepackage{xcolor}
\usepackage{multirow}

\usepackage{arydshln}
\usepackage{flushend}

\def\BibTeX{{\rm B\kern-.05em{\sc i\kern-.025em b}\kern-.08em
    T\kern-.1667em\lower.7ex\hbox{E}\kern-.125emX}}
\usepackage{enumitem}

\definecolor{brickred}{rgb}{0.8, 0.25, 0.33}

\usepackage{xspace}
\newcommand{\tool}{QF-Mixer\xspace}
\newcommand{\net}{QF-MixNN\xspace}
\usepackage{amsthm}
\newtheorem{principle}{Principle}
\theoremstyle{definition}
\begin{document}

\title{Exploration of Quantum Neural Architecture by Mixing Quantum Neuron Designs\\
\vspace{5pt}
% \Large (Invited Paper)
\vspace{-12pt}}
% Conference Paper Title*\\
\author{\IEEEauthorblockN{
Zhepeng Wang\textsuperscript{\dag},
Zhiding Liang\textsuperscript{\ddag}, 
Shanglin Zhou\textsuperscript{\S}, 
Caiwen Ding\textsuperscript{\S},
% Jinjun Xiong\textsuperscript{\P},
Yiyu Shi\textsuperscript{\ddag},
Weiwen Jiang\textsuperscript{\dag}}

\IEEEauthorblockA{\textsuperscript{\dag}George Mason University, VA, USA.
\textsuperscript{\ddag}University of Notre Dame, IN, USA.\\
\textsuperscript{\S}University of Connecticut, CT, USA.\\
% \textsuperscript{\P}University at Buffalo, NY, USA.\\
\{zwang48, wjiang8\}@gmu.edu
\vspace{-0.15in}}
}

\setlength{\textfloatsep}{2pt}
\setlength{\abovecaptionskip}{2pt}
\setlength{\dbltextfloatsep}{2pt}

% \author{\IEEEauthorblockN{1\textsuperscript{st} Zhepeng Wang}
% \IEEEauthorblockA{\textit{Dept. of Electrical and Computer Engineering} \\
% \textit{George Mason University}\\
% Fairfax, U.S. \\
% zwang48@gmu.edu}

% \and
% \IEEEauthorblockN{2\textsuperscript{nd} Zhiding Liang}

% \and
% \IEEEauthorblockN{3\textsuperscript{rd} Shanglin Zhou}
% \and
% \IEEEauthorblockN{4\textsuperscript{th} Caiwen Ding}

% \and
% \IEEEauthorblockN{5\textsuperscript{th} Jinjun Xiong}

% \and
% \IEEEauthorblockN{6\textsuperscript{th} Yiyu Shi}
% \IEEEauthorblockA{\textit{Dept. of Electrical and Computer Engineering} \\
% \textit{George Mason University}\\
% Fairfax, U.S. \\
% wjiang8@gmu.edu}

% \and
% \IEEEauthorblockN{7\textsuperscript{th} Weiwen Jiang}
% \IEEEauthorblockA{\textit{Dept. of Electrical and Computer Engineering} \\
% \textit{George Mason University}\\
% Fairfax, U.S. \\
% wjiang8@gmu.edu}
% }

\maketitle

\begin{abstract}
% With the consistent increase of quantum bit numbers in the actual quantum computers, implementing and accelerating the prevalent deep neural networks on quantum computers is becoming possible.
% Along with this, there emerges different designs of quantum neurons, and a fundamental question in quantum deep learning arises: what is the best quantum neural architecture?
% This paper makes the very first attempt to explore the quantum neural architectures, based on the existing neuron designs.
% We observe that the designs of quantum neurons are quite different, where the original variation quantum circuits can apply floating points for weights but can hardly be extended to multiple layers, while recent quantum neuron designs, like QuantumFlow, can build a multi-layer network, but has limitations on data representation (e.g., binary weights).
% To overcome these shortcomings, we propose to mixing different quantum neuron designs to take their respective advantages.
% In addition, we build an open-source tool\footnote{\todo{Accessing at https://github...}} for designers to easily integrate quantum designs and evaluate different combinations of basic quantum neurons in the exploration phase.
% Results demonstrate that the identified quantum neural architectures can achieve over xx\% of accuracy on 10-classes MNIST dataset, compared with xx\% on the state-of-the-art quantum neural network designs. 

With the constant increase of the number of quantum bits (qubits) in the actual quantum computers, implementing and accelerating the prevalent deep learning on quantum computers are becoming possible.
Along with this trend, there emerge quantum neural architectures based on different designs of quantum neurons. 
A fundamental question in quantum deep learning arises: what is the best quantum neural architecture? 
Inspired by the design of neural architectures for classical computing which typically employs multiple types of neurons, this paper makes the very first attempt to mix quantum neuron designs to build quantum neural architectures.
% through mixing the existing quantum neuron designs. 
We observe that the existing quantum neuron designs may be quite different but complementary, such as neurons from variational quantum circuits (VQC) and Quantumflow.
More specifically, VQC can apply real-valued weights but suffer from being extended to multiple layers, while QuantumFlow can build a multi-layer network efficiently, but is limited to use binary weights.
To take their respective advantages, we propose to mix them together and figure out a way to connect them seamlessly without additional costly measurement.
We further investigate the design principles to mix quantum neurons, which can provide guidance for quantum neural architecture exploration in the future. 
% In addition, we build an open-source tool\footnote{\todo{Accessing at https://github...}} for designers to easily integrate these quantum neuron designs in different combination ways and evaluate the corresponding quantum neural architectures in the exploration phase. 
Experimental results demonstrate that the identified quantum neural architectures with mixed quantum neurons can achieve 90.62\% of accuracy on the MNIST dataset, compared with 52.77\% and 69.92\% on the VQC and QuantumFlow, respectively.

% \todo{change abstract}
\end{abstract}


% \todo{Paper Title Change}

% \clearpage

\section{Introduction}
% This document is a model and instructions for \LaTeX.
% Please observe the conference page limits. 

% \begin{itemize}
%     \item Background on scaling up of actual quantum computers
    
%     \item Background of quantum neuron designs.
%     \item Quantum neural network design with the observation on different quantum neurons.
%     \item Our contributions: (1) mixing neurons; (2) evaluation tool.
%     \item three-fold contributions
%     \item results
%     \item paper roadmap
% \end{itemize}

% \wwnote{[Background]}

Neural networks have shown its great power on machine learning applications such as image classification~\cite{he2016deep}, speech recognition~\cite{graves2013speech} and visual question answering~\cite{antol2015vqa}. 
With the growing demand on high accuracy, the size of neural networks consistently grow: from $60$ million parameters in the very first image classification task model (i.e., AlexNet~\cite{krizhevsky2017imagenet} to over $175$ billion parameters in the latest speech recognition model (i.e., GPT-3~\cite{brown2020language}).
The growing model size gradually sets a gap between the requirements/demands and supplies in classical computing.
Quantum computing~\cite{bertels2021quantum,jiang2021machine,burgholzer2020advanced}, on the other hand, is evolving rapidly in recent years: from 5 quantum bits (qubits) back to 2016 to 65 qubits in 2020, and IBM plans to develop a quantum processor with 1,121 qubits in 2023~\cite{ibm2020roadmap}.
In consequence, there emerge recent works on putting neural computation to quantum computing~\cite{yan2020nonlinear,tacchino2019artificial,jiang2021co,sim2019expressibility,schuld2021effect,romero2021variational,lockwood2020reinforcement,benedetti2019parameterized,khairy2020learning}, and it is expected that neural networks on quantum computers (or quantum neural networks, QNNs) can achieve exponentially speedup, compared with the classical version.

% with much higher parallelism of quantum accelerator using quantum computing makes it the most promising solution compared with the other accelerators like ASIC, FPGA based on classical computing~\wwnote{[cite]}. The bottleneck that hinders the quantum computers being applied to large-scale computation is its limited number of quantum bits (qubits). However, this number has been increased constantly and rapidly in the last few years and is expected to grow to 1121 qubits reported by IBM~\wwnote{[cite]}.

% In a neural network, t

When the neural network comes to quantum computing, a fundamental question arises: what is the best neural architecture for quantum computing and can we directly apply the ones designed for classical computing?
Limited by the computation can be performed and the data can be retained by the quantum computer, not all classical neural operations can be directly applied to quantum computing.
For example, N qubit can represent $2^N$ data items, but their value is limited by the range from -1 to 1.
As a result, directly applying the neural network architecture designed for classical computing cannot work.

% Neural architecture is constructed based on the artificial neuron. To understand what is the best neural architecture, different quantum version of artificial neuron (a.k.a., quantum neuron) designs were proposed recently.
% For example, quantum perceptron neurons are proposed by \cite{tacchino2019artificial,jiang2021co,jiang2021machine}, which can utilize quantum computing to realize the perceptron.
% Authors in \cite{de2019implementing,yan2020nonlinear} proposed the quantum non-linear neurons based on the boolean functions, so that non-linear function (e.g., ReLU) can be realized.
% In \cite{jiang2021co}, authors proposed the quantum normalization neuron to adjust the outputs of quantum neurons.
% In addition to these implementation having analogue on classical computing, variational quantum circuit (VQC) \wwnote{[cite]} is proposed to be a basic neuron to perform machine learning tasks, which utilizes the entanglement provided by quantum computing to realize highly correlated neurons that are difficult to be implemented on classical computers.

In recent years, different quantum neurons (e.g., quantum perceptron, quantum non-linear neurons, variational quantum circuit) are proposed \cite{de2019implementing,yan2020nonlinear,tacchino2019artificial,jiang2021co,schuld2020circuit,sim2019expressibility,schuld2021effect,romero2021variational,chen2020variational,schuld2019quantum,lockwood2020reinforcement,benedetti2019parameterized,khairy2020learning}.
On top of the basic quantum neurons, the architecture of QNN needs to be explored for complicated machine learning tasks.
Although the QNNs built with the proposed quantum neurons achieve good results on a small dataset, for example, over 90\% accuracy on a subset of MINST (e.g., only 3 and 6 digits), they will perform much worse for a larger dataset (say 52.77\% on a full MNIST dataset using VQC).
It is, therefore, at most of importance to optimize the neural network architecture for quantum computing.

The experience from the neural architecture design~\cite{huang2017densely,jiang2019achieving, howard2017mobilenets, song2021dancing,mehta2019espnetv2, wu2020intermittent,zhang2018shufflenet, wang2021lightweight,iandola2016squeezenet, jia2020personalized} on classical computers can shed light on the design of QNN to improve performance. 
One way in the design of classical NN is to mix the basic neurons in existing networks to explore a better neural architecture, which is typically used in neural architecture search (NAS)~\cite{cai2018proxylessnas,jiang2020standing,liu2018darts,jiang2020device,wu2019fbnet,jiang2020hardware,zoph2016neural,yang2020co,tan2019mnasnet,yang2020co2,jiang2019accuracy}.
For example, the dilated depthwise-separable convolution from ESPNet~\cite{mehta2019espnetv2} and depthwise-separable convolution from MobileNets~\cite{howard2017mobilenets} are commonly used in NAS.
With the exiting designs of quantum neurons, it seems straightforward to mix them together for QNN architecture exploration; however, new questions and challenges are posed in connecting different quantum neurons: it is unclear (1) whether different quantum neuron designs can be complementary to each other by mixing them, instead of being canceled out by others;
(2) whether we can connect any types of quantum neurons, considering that quantum neurons may have different data encoding; and (3) whether we can avoid use measurement operation at the joint point of quantum neurons, since the measurement incurs classical to quantum communication which may easily become a performance bottleneck.

In this paper, we make the very first attempt to investigate the potential of mixing different quantum neurons.
More specifically, we develop the design principles of mixing quantum neurons, which can be applied to a wide range of quantum neuron designs using different data encoding methods.
This will be a guidance for the future exploration of quantum neural architectures using different quantum neurons and connecting them seamlessly. 
Follow the developed design principles, we identify a mixed neural architecture, namely \net, which incorporates both VQC~\cite{sim2019expressibility} and neurons in QuantumFlow \cite{jiang2021co} such that they can be complementary for each other and no measurement is needed during the execution.
To evaluate the proposed \net in terms of network performance (e.g., accuracy) and the building cost, we further provide an exploration tool for connecting quantum neurons, namely \tool.

% we explore the potential of following the second method to get the best quantum neural architecture. However, mixing the existing quantum neurons is not a trivial problem. First, the design philosophy/their characteristic is conflict, which could make the advantage brought by a specific neurons be canceled out by the other mixed neurons. Second, due to the diversity of encoding method of quantum neurons, the connections of some pair of different neurons might be infeasible or could cause large overhead on training. Considering these major challenges, we figured out that the quantum neurons from VQC-based QNN and Quantumflow are different but complimentary~\wwnote{[this point will be illustrated in motivation part]}, which is suitable to be mixed. Moreover, by carefully considering all the possible connections between different neurons, we find a method to connect these different neurons seamlessly without additional costly measurement. 

% \wwnote{[Contribution]}

The contribution of this paper is three-fold as follows. 
\begin{itemize}
    \item We investigate the connection of quantum neurons with different encoding methods and propose the design principles for mixing quantum neurons. It gives the insight and guidance for mixing different quantum neurons.
    \item We seamlessly connected by VQC-based QNN and neurons in Quantumflow to construct a new quantum neural network, \net, which can outperform the existing QNNs.
    % This is the first work to design the quantum neural architecture to mix the different neurons from VQC-based QNN and Quantumflow.
    \item \tool\footnote{{Accessing at https://github.com/Jqub/QF-Mixer}} is developed for designers to easily mix the quantum neuron designs in different ways and evaluate the produced quantum neural architecture during the exploration phase.
    % \todo{use which link?}
    \item Experimental results show that \net can achieve 90.62\% of accuracy on MNIST, compared with 52.77\% and 69.92\% on VQC and QuantumFlow, respectively.
\end{itemize}

% \wwnote{[Results]}

% \todo{double check results before finalizing}
% \todo{Is it ok to be the same as abstract?}

% Experimental results show that our proposed quantum neural architectures can achieve 90.62\% of accuracy on MNIST dataset, compared with 52.77\% and 69.92\% on the VQC and QuantumFlow, respectively.

% \wwnote{[Paper roadmap]}

% \todo{double check roadmap before finalizing}

%structure of the paper
The remainder of the paper is organized as follows. Section~\ref{Sec:Priliminary} reviews the detailed design of neurons from VQC-based QNN and QuantumFlow and presents the motivation of mixing them for quantum architecture design. Section~\ref{Sec:Method} provides a detailed description of our proposed quantum neural architecture. Section~\ref{Sec:Theory} discusses all the possible connection between different neurons and points out the critical principles to follow for the connection.  Experimental results are shown in Section~\ref{Sec:Experiment} and the  concluding remarks are given in Section~\ref{Sec:conclusion}.

% \subsection{Variational (Parameterized) Quantum Circuits}

% \subsection{Quantum Neuron}
% \wwnote{This mean the set of quantum neuron: like Q-Neuron, U-NEU and P-NEU in QuantumFlow.}

% \subsection{Quantum Deep Neural Network}
% \todo{Seems not deep, hybrid QNN}
% \subsection{Role of Non-Linearity in Deep Neural Network}
% \todo{We could mention it when we talk about the limitation of VQC. Universal approximation, summer school slides. (Day 5?)}

% \subsection{My Outline}
% \begin{enumerate}
%     \item Background introduction from quantum computing to quantum machine learning, Distinguish QVC. List the advantage and limitation?
%     ~\todo{Refer to Dr. Jiang's slides, summer school slides and lab}
%     \item Brief introduction of existing DNN. Distinguish the quantumflow. List the advantage and limitation? \todo{Read quantumflow paper}
%     \item Briefly summarize the advantage and disadvantage of QVC and QF. Emphasize the challenge.
%     \item Propose the motivation of our paper, i.e., The combination. Illustrate how the combination overcome the separate weakness or solve the challenges.
%     \item Briefly summarize the contribution of this paper. We can also add some brief description of the experimental results here to further demonstrate our claim. 
%     \item Briefly conclude the organization of the following sections. 
    
% \end{enumerate}

% \clearpage
\section{Preliminary and Motivation}\label{Sec:Priliminary}\label{Subsec:VQC}\label{Sec2:QF}\label{Subsec:Motivation}

% % \subsection{Background}
% \wwnote{Add discussions with all neuron designs in the slides.}

\begin{figure}[t]
\centering
\includegraphics[width=0.99\linewidth]{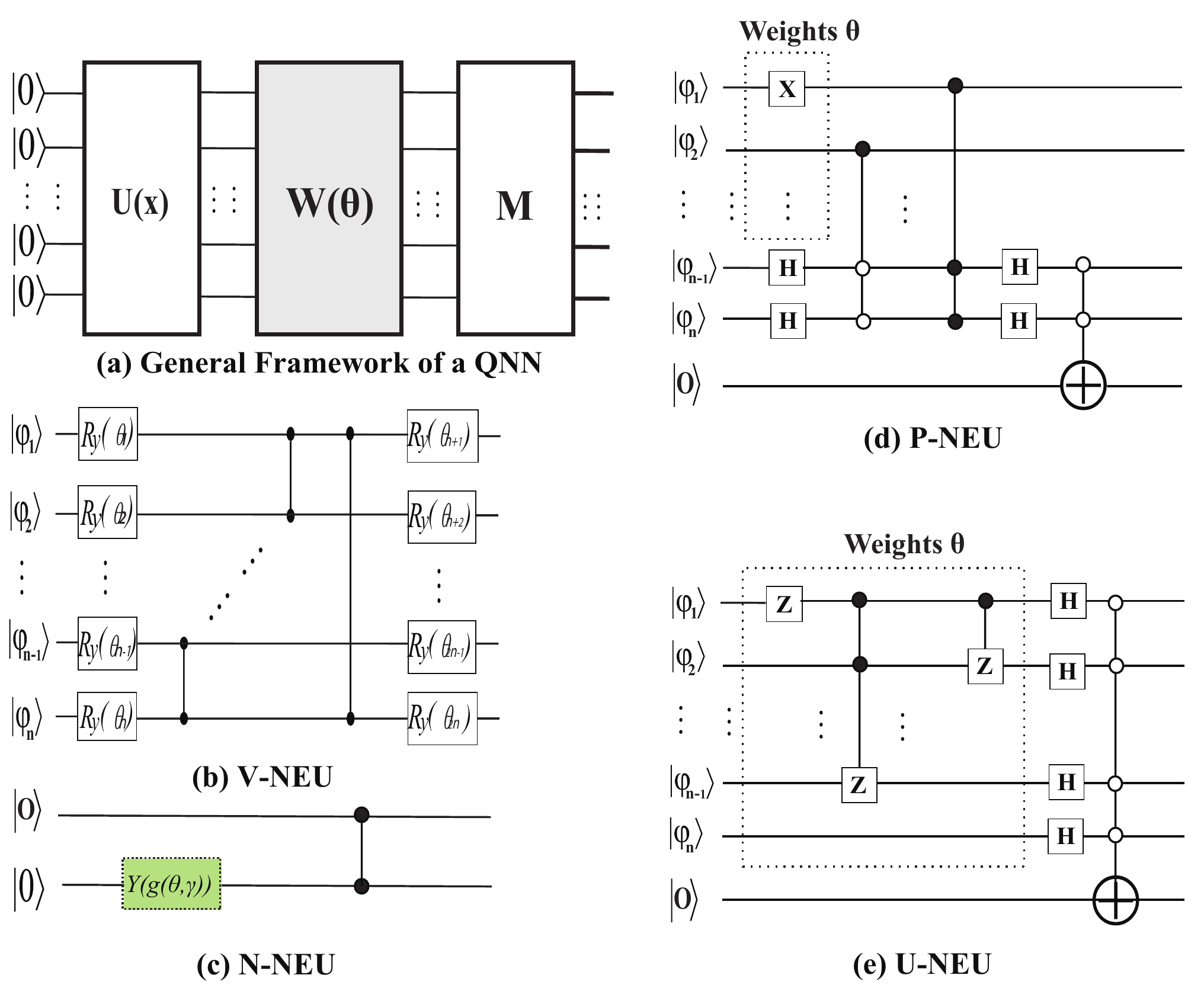}
\caption{Illustration of QNN and quantum neurons: (a) general framework of a QNN, where $\theta$ is the adaptable parameters of the QNN; (b) V-NEU with $2n$ adaptable parameters and $n$ input qubits, which is reused as $n$ output qubits; (c) N-NEU for normalization; (d) P-NEU with $n$ input qubits and a single output qubit; (e) U-NEU with $n$ input qubits and a single output qubit.}
\label{fig:motivation}
\end{figure}

Neural architecture is constructed based on artificial neurons. To understand what is the best neural architecture, different quantum versions of artificial neuron (a.k.a., quantum neuron) designs were proposed recently.
For example, quantum perceptron neurons are proposed by \cite{tacchino2019artificial,jiang2021co,jiang2021machine}, which can utilize quantum computing to realize the perceptron.
Authors in \cite{de2019implementing,yan2020nonlinear} proposed the quantum non-linear neurons based on the boolean functions, so that non-linear function (e.g., ReLU) can be realized.
In \cite{jiang2021co}, the authors proposed the quantum normalization neuron to adjust the outputs of quantum neurons.
In addition to these implementations having analogue on classical computing, variational quantum circuit (VQC) \cite{schuld2020circuit,sim2019expressibility,schuld2021effect,romero2021variational,chen2020variational,schuld2019quantum,lockwood2020reinforcement,benedetti2019parameterized,khairy2020learning} is proposed to be a basic neuron to perform machine learning tasks, which utilizes the entanglement provided by quantum computing to realize highly correlated neurons that are difficult to be implemented on classical computers.

{Among all the above quantum neuron designs, VQC is widely used in different neural networks, like Quantum GAN~\cite{dallaire2018quantum}, Quantum LSTM~\cite{chen2020quantum}, etc. On the other hand, QuantumFlow is the very first work to demonstrate the potential quantum advantage that can be achieved on quantum circuits.
In the following, we will provide a detailed introduction to these two designs, both of which will also be the basis of our proposed neural network design with mixed quantum neurons.}

\vspace{5pt}
\noindent\textit{A. {VQC-Based Quantum Neural Network}}
\vspace{5pt}
% Variational quantum circuit (VQC), also known as parameterized quantum circuit, is a type of quantum circuits which include quantum gates with \textit{real-valued} adaptable parameters. With appropriate optimization algorithm, VQC could get the best configuration of its parameters to achieve the goal of targeted task compared with a standard quantum circuit without any adaptable parameters. Therefore, VQC has been widely used in applications on quantum chemistry, chemical engineering and so on. 
% Since most of machine learning tasks could be formulated as an optimization problem, it is natural to use VQC as the model to learn from dataset and then be applied to specific tasks. 

% In recent years, neural networks (NN) have become the most powerful model to solve the machine learning problem. Therefore, 

VQC-based quantum neural network~\cite{sim2019expressibility} is built up to take use of the entanglement from quantum computing to model complex and highly correlated distributions, expecting to achieve higher performance over the classical neural network.

% typical QVC-based QNN model.
% A VQC-based QNN

Figure~\ref{fig:motivation}(a) shows the general framework of a QNN, which mainly consists of three components: (1) a series of quantum gates with non-adaptable parameters for state preparation; {(2) a series of quantum gates with adaptable parameters $\boldsymbol{\theta}$ to mimic biological neurons;} (3) a set of measurement operations to extract output.

The VQC-based QNN follows the general framework in Figure~\ref{fig:motivation}(a). The key design of it is the second component, W($\boldsymbol{\theta}$). Figure~\ref{fig:motivation}(b) is an example of a typical and basic VQC. And W($\boldsymbol{\theta}$) is built by repeating such basic VQC multiple times. As shown in Figure~\ref{fig:motivation}(b), the VQC with $n$ input qubits contains $2n$ Ry gates with $2n$ adaptable \textit{real-valued} parameters $\theta_i (i = 1,2,...,2n)$. Note that the $n$ input qubits are reused as $n$ output qubits. 
The repetition of such VQC could increase the number of adaptable parameters and thus improve the model's capacity. By tuning the value of $\theta_i$, the model can be optimized for the given tasks.

% the basic circuit block $U(\boldsymbol{\theta})$ for $n$ times. 

% its optimal configuration for the given task could be obtained. 

% The first component is a quantum circuit for state preparation.
% It contains non-adaptable parameters, taking the classical data $\boldsymbol x$ as input and transforming the zero state to the corresponding quantum state $|x\rangle$. 

% The third component measures the output qubits of VQC to extract the outputs and post-process will be conducted to obtain classical results (e.g., a class for a classification problem). 

% Quantum Computers can naturally model complex, highly correlated distributions
% developed for machine learning tasks.
% to take use of the entanglement from quantum computing for .

% while taking the efficiency advantage of quantum computing. 

% It 

Although the VQC-based QNNs have been widely applied to machine learning tasks~\cite{sim2019expressibility}, it lacks the universal approximability because quantum gates are intrinsically linear operations, which are difficult to construct non-linear operations  within the quantum circuits. 
Therefore, it could not handle the dataset that is not linearly separable. For example, this kind of QNN could not approximate an XOR operation, whose input is not linearly separable.

Inserting the measurement components between the basic circuit blocks (e.g., VQC in Figure~\ref{fig:motivation}(b)) within the VQC-based QNNs is a potential solution to implement the non-linear function (i.e., quadratic function).
However, it will 
% Note that although the measurement component implements a non-linear function (i.e., quadratic function) implicitly, it will 
bring the communication overhead between the quantum-classical interface during the computation.
Such overhead can easily dominate the overall performance, and cancel out the benefit brought by quantum computing further.

% if we add it to connect two basic circuit block, which might cancel out the benefit brought by quantum computing. 

An alternative solution to achieve universal approximability is to add extra neural network layers which will be executed on the classical computer.
Similar to the previous solution, we cannot avoid the quantum-classical communication overhead. Moreover, this solution will offload part of neural computations to classical computers, which mitigates the advantages brought by quantum computing.

% for VQC-based QNN
% Some works also propose to use an extra NN taking the output data from the VQC-based QNN as input and run the extra NN on the classical computer to solve the lack of universal approximability problem. But again, this method still incur the quantum-classical communication overhead and it also lowers the proportion of quantum part in the whole NN, which mitigates the positive effect of quantum computing.
\vspace{5pt}
\noindent\textit{B. {QuantumFlow}}
\vspace{5pt}

Quantumflow\cite{jiang2021co} is a co-design framework to optimize quantum neurons and quantum circuits simultaneously, such that quantum advantage can be achieved.
In terms of different quantum circuit implementations, two 
% is the first work to co-design the architecture of QNN and their corresponding quantum circuit implementations. To achieve a simple and efficient circuit implementation, the authors customize two 
types of quantum neurons were proposed, including
% for the neural computation of QNNs, i.e., 
P-NEU and U-NEU.

\textbf{P-NEU} implements a probabilistic model-based neural computation. For P-NEU, the state-preparation component should encode the original \textit{real-valued} input data to the quantum states $|\phi_{i}\rangle (i = 1,2,...,n)$ through \textit{probability encoding}, which is then used as the input for P-NEU. More specifically, each data item will be encoded to the probability of a qubit's $|1\rangle$ state, which will be illustrated in Section~\ref{Sec:Theory}. P-NEU embeds the adaptable weights $\theta$ to the input qubits by using X gates (denoted by the dashed box in Figure~\ref{fig:motivation}(d)), followed by a circuit to complete the weighted sum operation of inputs through Hadamard gates and Control-X gates as shown in Figure~\ref{fig:motivation}(d). The single output qubit $|O\rangle$ will be measured to extract the output data and perform post-process, which corresponds to $M$ in Figure~\ref{fig:motivation}(a).

% ~\todo{P-NEU neural computation figure; Add $w_k$ to the neural figures}

% Figure~\wwnote{[ref]} shows the neural computation of P-NEU. 
% The R operation converts the real-valued input feature $I_k$ that ranges from 0 to 1, to a quantum state $|x_k\rangle$ of quit $k$, where $|x_k\rangle=\sqrt{1-p_k^2}|0\rangle+\sqrt{p_k}|1\rangle$, which implements the encoding of feature $I_k$. Mathematically, the encoded quantum state could represent a two-point distributed random variable $x_k$, where $P\{x_k=-1\}=p_k$ and $P\{x_k=+1\}=1-p_k$. The C operation calculates the weighted average sum of inputs, where the weight $w_k$ is adaptable parameter. The A operation is nonlinear (i.e., quadratic) activation function which output $y^2$ as a random variable. Finally, The E operation calculates the expectation of the random variable $y^2$ and thus outputs a 0–1 real number. 

% \todo{add U-NEU nerual computation}

% Different from P-NEU , 

\textbf{U-NEU} is based on the input qubits encoded with \textit{amplitude encoding} (see details in Section~\ref{Sec:Theory}), which maps an input data item to the amplitude of a basis state (e.g., $|0\rangle$ or $|1\rangle$ for one qubit). As a result, for a quantum computer with $k$ qubits, it can encode $2^k$ input data items. U-NEU encodes the adaptable weights $\theta$ to the input qubits by using Z gate and Control-Z gates (denoted by the dashed box in Figure~\ref{fig:motivation}(e)).
Next, similar to P-NEU, we apply a circuit to complete the weighted sum operation of inputs through Hadamard gates and Control-X gates and store the output to an ancillary qubit.
Last, the output qubit $|O\rangle$ is measured and we perform post-process on the output, which corresponds to $M$ in Figure~\ref{fig:motivation}(a).

% is a quantum neuron based on the unitary matrix, which will first encode the $n = 2^k$ inputs to the amplitudes of states in a quantum circuit with $k$ qubits, through singular value decomposition (SVD) to convert the input to the first column vector in the unitary matrix. This type of encoding method is called \textit{amplitude encoding}. 

% Figure~\wwnote{[ref]} shows the neural computation of U-NEU. The quantum circuit represented by $U$ implements the amplitude encoding of input. The following steps of the neural computation is similar to the process of P-NEU. But for the C operation, y equals to the weighted sum of inputs divided by $\sqrt{m}$ instead of $m$. And since U-NEU is not probabilistic model based, the A operation calculates $y^2$ directly without the need to calculate its expectation. 

% \todo{Add the circuit figure of U-NEU and P-NEU} 

In addition to P-NEU and U-NEU, QuantumFlow further devises N-NEU for normalization, shown in Figure~\ref{fig:motivation}(c).
% Figures~\wwnote{[ref]}, ~\wwnote{[ref]}, and \wwnote{[ref]} illustrate the quantum circuit implementation of these 3 layers. 

In the circuit implementations of P-NEU and U-NEU, the multiplication between inputs (i.e., $x_k$ for P-NEU and $u_k$ for U-NEU) and weights (i.e., $w_k$) are realized by maintaining or flipping amplitudes of a quantum state.
% Both of them 
% utilize the phase flip to realize the multiplication between input $x_k$ (or $u_k$ for U-NEU) and $w_k$. 
\textbf{As a result, adaptable weight $\theta$ has to be \textit{binary}}, which degrades the representation capability of the model compared with \textit{real-valued} weight.

{The networks devised by QuantumFlow (referred as QF-Net) apply the above three neurons, which shows the potential advantage to have different quantum neurons when designing a QNN; however, both neurons have the same limitation on data representation, i.e., binary weights.
In this work, we aim to provide a more general framework to incorporate different neuron designs to form a QNN. And we investigate the requirements and overhead in connecting different neurons.}

% Benefiting from the quantum-aware data interpretation for inputs, P-NEU can be attached to the output qubits of previous layers without measurement;
%  Each feature value of a input data is modeled as random variables following tow-point distribution, which could then be easily encoded as a qubit in superposition form.

% the pure quantum computing based neural
% computation. the pure quantum computing
% based accelerator; pure quantum computing; pure quantum computing design

\vspace{5pt}
\noindent\textit{C. {Motivation: VQC and QuantumFlow are complementary}}
\vspace{5pt}

We observe that VQC-based QNN and QF-Net are complementary to each other, and it is meaningful to mix them in a holistic neural network for better neural model performance.
% Details are discussed as follows. 

\textbf{VQC-based QNN enhances QF-Nets}. The weights of VQC-based QNN are \textit{real numbers} while the weights of QF-Net are restricted to \textit{binary numbers}. It is obvious that real-valued weights have higher representation ability since binary-valued weights are just the special case of the real-number ones.
% Although DNN built upon binary weight, i.e., binarized neural network (BNN), has been proved to have the ability to achieve relatively high accuracy in the deep learning applications~\cite{courbariaux2015binaryconnect}, 
% Applying real numbers to the weights could significantly expand the search space in the training phase, compared with the neural network applies binary weights. 
Thus, there exists great potential to improve the model performance (i.e., accuracy for classification tasks) by taking VQC as a complement to QF-Net.

% could enable the QF-Net to have real-valued weights and thus makes the QF-Net more powerful. 

\textbf{QF-Nets enhances VQC-based QNN.} VQC-based QNN has difficulties in integrating non-linear function into the quantum circuit without measurement, which may not have universal approximability, resulting in low neural model performance.
On the other hand, QF-Nets provide flexibility in realizing quadratic function in-between layers.
In consequence, mixing the VQC and QF-Net provide the potential to make the network having universal approximability, and in turn, improve model performance.

\begin{figure*}[t]
\centering
\includegraphics[width=0.99\linewidth]{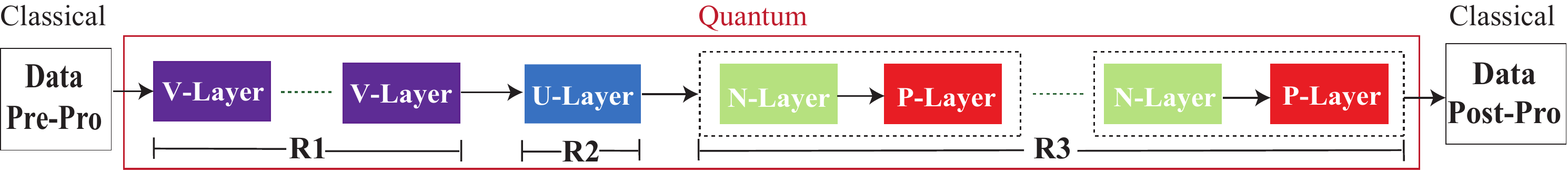}
\caption{Overview of the structure of \net.}
\label{fig:method}
\end{figure*}

\textbf{Challenge.} It seems straightforward to mix different quantum neurons to build up a network; however, arbitrary connecting different neuron designs may bring heavy overheads, posing new challenges.
More specifically, if measurement operation is inserted at the joint point of quantum neurons, it can become the performance bottleneck and diminish the benefits brought by quantum computing.
However, it is unclear whether or not we can avoid use measurement operation in a neural network with mixed quantum neurons.

\section{\net: A Quantum Neural Architecture with Mixed Quantum Neurons}\label{Sec:Method}
% \todo{Is the title good?}

% \todo{Illustration of encoding?}

Although connecting different quantum neurons arbitrarily seems feasible when they are implemented on quantum circuits, mixing them without any rules could cause problems that make the built quantum architecture difficult to be used in practice, due to high cost and not scalable.
% with low cost or be scaled to the large input size. 
To make the quantum neural architecture cost efficient and scalable, there are several goals needed to be achieved as follows.

\textbf{Goal 1:  The entire quantum neural architecture is able to be executed on quantum processor without measurements.}
In this way, we can avoid the expensive quantum-classical communication overheads.
% , which is mentioned in Section~\ref{Sec:Priliminary}.

\textbf{Goal 2: The same type of neuron should be consistent regardless of the location of the neuron.} For example, if a neuron as the first layer applies amplitude encoding to the input data. When this neuron is used in the other layers, its encoding method for input should still be amplitude encoding. In this way, we can build neural blocks to support QNN.

\textbf{Goal 3: The quantum architecture could be trained on classical computers correctly and efficiently.} 
% Quantum processors in NISQ era have limited number of qubits and high noise, directly training QNN on quantum processors is unstable and not scalable, and therefore, it requires the assistance of classical computer for training.
% Therefore, we have to train it on classical computers. A straightforward way for such training is to formulate each quantum gate as a unitary matrix. However, since most of quantum neurons 
% Changing it to probability encoding is not allowed. 
% ~\wwnote{[do we need to explain why?]}
% \todo{explain point 2 here?}
Although training DNN on classical computers is expensive and inefficient~\cite{wu2020enabling,wang2019e2,wu2021enabling}, we still have to train QNN on classical computers because the near-term quantum computer (NISQ) has a limited number of qubits and high noise on each qubit, resulting in the QNN training on quantum computers being unstable and not scalable. A straightforward way to achieve the equivalent training on classical computers is to formulate each quantum gate as a unitary matrix. However, since most of layers apply quantum gates that entangle the input qubits, \textbf{the output qubits of each layer are coupled with the output qubits of all the preceding layers}. For example, assume that layer $i$ has $p$ output qubits and the number of output qubits from all of the previous layers are summed to $m$. Then the size of unitary matrix to formulate the quantum gate in layer $(i+1)$ is $O(2^{m+p}) \times O(2^{m+p})$. It is obvious that when the QNN goes deeper, the size of the unitary matrix for the last layer of the QNN will be increased dramatically, which incurs expensive training overhead on memory and computation. Fortunately, the output qubits of layer $i$ could be regarded as decoupled with the output qubits of previous layers in some cases, which will be illustrated in Section~\ref{Sec:Theory}. If this is the case, then the size of the unitary matrix for the quantum gate in layer $i+1$ will be decreased to $ O(2^{p}) \times O(2^{p})$, which is only related to the number of output neurons of the current layer and thus reduces the training overhead by a large margin. Besides, the first and third goals are also naturally achieved if mixture is conducted between the quantum neurons of layer $i$ which is decoupled with its preceding layers, and the quantum neurons of layer $i+1$.

% \todo{double check the figure description，Add R1 to U, preprocess and postprocess}

To achieve all of the above goals, we propose \net, a novel design mixing VQC (called V-NEU in the following for consistency), U-NEU, P-NEU, and N-NEU together. 
Figure~\ref{fig:method} shows the structure of \net:
After data preprocessing, an input data with $N$ data items will be encoded as the corresponding quantum state with $\log N$ qubits through \textit{amplitude encoding}. It will be sent to V-Layer, which is implemented by V-NEU. V-Layer could be repeated by $R_1$ times, where $R_1 \ge 1$. By increasing $R_1$, the number of real-valued weights are increased. 

Then, $R_2$ of U-Layer $(R_2 \in \{0,1\})$ will be added. Note that if $R_2=0$, we skip U-Layer and connect V-Layer with other layers to add flexibility. The U-Layer will take the output from the previous V-Layer as input, execute the neural computations implemented by U-NEU, and output the quantum state encoded as probability. The next block is made up of N-Layer and P-Layer, where N-Layer and P-Layer consist of multiple N-NEUs and P-NEUs, respectively. Such a block could be repeated by $R_3$ times, where $R_3 \ge 0$. The quantum states produced by the last layer of \net will be measured in the data post-processing stage and the quantum state of each output qubit will be interpreted as the probability of its corresponding class, which is used for classification. 

Assume $N_i$ denotes the number of output qubits of the $i$- th layer. If the layer is V-Layer, since V-NEU reuses the input qubits as its output qubits, $N_i = N_{i-1} = ... = N_{1} = \log N$. For U-Layer and P-Layer, $N_i$ equals the number of neurons (i.e., U-NEU and P-NEU, respectively), which is manually defined.
% and could be tuned for higher accuracy with our future work using neural architecture search (NAS) to find the optimal configuration. 
For N-Layer, since it always follows either U-Layer or P-Layer, $N_i$ should be the same as the number of neurons of its previous layer, i.e., $N_{i-1}$. Note that for the last layer in \net, its number of neurons has to be set to the number of classes defined by the classification tasks regardless of the type of its neurons. 

\section{\tool: Design Principles}\label{Sec:Theory}
% \todo{Title?}

\tool is developed to help designers in exploring different quantum neural architectures.
The key component in \tool is to check whether or not two types of quantum neurons can be connected without violating the design goals presented in Section \ref{Sec:Method}; if yes, what is the cost?
The above question is non-trivial, because the neuron designs may apply different data encoding methods.
In this section, we will investigate two most commonly used data encoding methods: (1) amplitude encoding, and (2) probability encoding, a.k.a., angle encoding.
We will provide 5 design principles to cover all possible situations for mixing quantum neurons applying the above encoding methods.

Before introducing the design principles, we first introduce two data encoding methods.
\textit{Amplitude encoding (A)} is to encode $N$ data items to $N$ amplitudes of quantum state composed of $\log_2N$ qubits.
For example, the given dataset is $\{a,b,c,d\}$, amplitude encoding will operate 2 qubits $|q_0 q_1\rangle$, such that $|q_0 q_1\rangle=a|00\rangle+b|01\rangle+c|10\rangle+d|11\rangle$.
\textit{Probability encoding (P)} is to interpret the given data $d$ as a probability and apply a rotation gate (e.g., $Ry$ gate) with an angle $\theta$ on a qubit (say $|\psi\rangle=|0\rangle$), such that the qubit has the probability of $d$ to be state $|1\rangle$, i.e.,
$Ry|\psi\rangle=\sqrt{1-d^2}|0\rangle+\sqrt{d}|1\rangle$.
To better study the connection of different neurons, we give the data encoding method for different neuron designs. Note that the input and output may apply different encoding methods. 
\begin{itemize}[noitemsep,topsep=0pt,parsep=0pt,partopsep=0pt]
    \item U-NEU: Input (A) and Output (P).
    \item V-NEU: Input (A) and Output (A/P).
    \item P-NEU: Input (P) and Output (P).
    \item N-NEU: Input (P) and Output (P).
\end{itemize}
As we can see from the above list, we need to investigate all possible combinations of data encoding, which helps us to formulate the design principles.

\begin{figure}[t]
\centering
\includegraphics[width=0.8\linewidth]{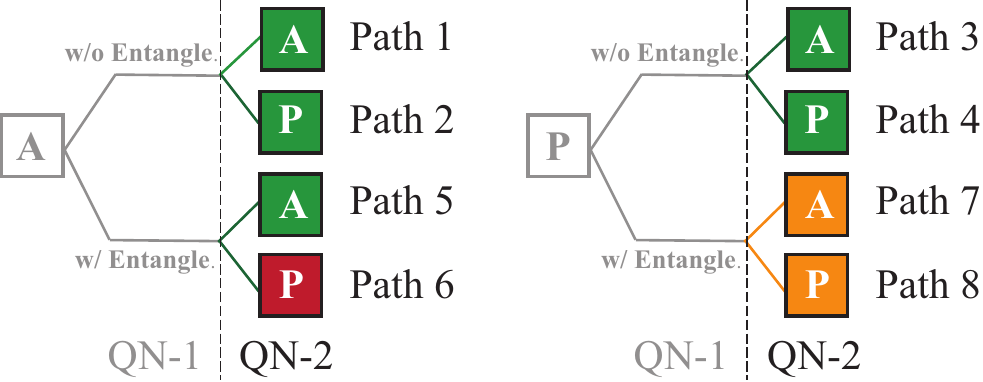}
\caption{8 possible combinations of data encoding for one neuron (QN-1) and its descendant (QN-2). (Best viewed in color)}
\label{fig:principle}
\end{figure}
 
% \todo{Add general description of the figure. And also double check the when the figure is finalized}

Figure~\ref{fig:principle} shows all 8 combinations/paths when two different neurons are connected. 
Specifically, we have two separate graphs since the output of the neurons in the first layer (denoted as QN-1) can be either $P$ or $A$ encoding.
For each case, there are two paths in terms of whether or not the output qubits are entangled after the neuron computation in QN-1. 
Furthermore, the input of neurons in the second layer (denoted as QN-2) has two possible encoding methods.
As a result, there are 8 paths in total.

Given two types of neurons, \tool not only needs to identify the path for these neurons, but also need to tell whether such path is feasible.
Here, a path is feasible, if a combination can achieve the design goals mentioned in Section~\ref{Sec:Method}.
In the paper, the paths that are feasible if the node in QN-2 is marked in green, and it is not feasible, if it is in red, and it is conditional feasible for that in orange color.

% And if a combination could not achieve the deign goals mentioned in Section~\ref{Sec:Method}, it is regarded as infeasible. Each path in Figure~\wwnote{[ref]} corresponds to one possible connection. The path with the last branch in red (blue) represents its corresponding connection is infeasible (feasible). When the last branch is in yellow, the corresponding connection is feasible only when it satisfy the specific conditions. For instance, Path 6 in Figure~\wwnote{[ref]} means that if the output qubits from the first quantum neuron (QN-1) is entangled (i.e., incoherent) and encoded as amplitude while the the second quantum neuron (QN-2) also expects its input to be encoded as amplitude, then connecting these two neurons is feasible.

% \todo{expectation of input}

% \wwnote{[List the principle here]}

% Based on Figure~\wwnote{[ref]}, the critical principle for mixing different quantum neurons will be illustrated as follows.\\\\

\tool figures out 5 design principles, listed as follows.

\begin{principle}\label{p:path3478}
\textbf{(Paths 1-4)} If the output qubits of QN-1 are not entangled, then it is feasible to be directly connected to QN-2 regardless of the encoding method of QN-2's input. 
\end{principle}

Without quantum entanglement, the output qubits from QN-1 are decoupled with the output qubits of its previous layers. Therefore, the direct connection between QN-1 and QN-2 is feasible.
In the following, we will discuss the situation that the output qubits of QN-1 are entangled.

% Path 3, 4, 7 and 8 in Figure~\wwnote{[ref]} correspond to Principle 1.\\\\

\begin{principle}\label{p:path6}
\textbf{(Path 5)} 
If output qubits of QN-1 are amplitude encoding and entangled, and the input of QN-2 is amplitude encoding, then it is feasible to connect QN-1 and QN-2.
\end{principle}

\begin{principle}\label{p:path5}
\textbf{(Path 6)} 
If output qubits of QN-1 are amplitude encoding and entangled, and the input of QN-2 is probability encoding and having independent requirement, then it is infeasible to connect QN-1 and QN-2.
% If QN-1 contains quantum gates that entangle its output qubits and its output qubits use \textit{amplitude encoding}, then it is infeasible to connect it directly to QN-2 when the expected encoding method of the input of QN-2 is \textit{probability encoding}.
\end{principle}

The neuron designed using probability encoding commonly assume that the input data are independent, so that the rotation gate can be applied for the angle encoding. According to the third design goal in Section~\ref{Sec:Method}, QN-2 also has such requirement. However, the output qubits of QN-1 are entangled, which violates such requirement. Thus, we have the above principle.
% When no probability encoding is involved in the encoding of output qubits of QN-1 and input qubits of QN-2, the design goals could be achieved regardless of whether QN-1 and QN-2 are decoupled. 
% \wwnote{The above explanation seems meaningless?}
This principle is obvious.

\begin{principle}\label{p:path2}
\textbf{(Path 7)} 
If output qubits of QN-1 are probability encoding and entangled, and the input of QN-2 is amplitude encoding, then it is feasible to directly connect QN-1 and QN-2 if the output qubits of QN-1 are the same with its input qubits.
% whether the direct connection between QN-1 and QN-2 is feasible is conditional when the expected encoding method of the input of QN-2 is \textit{amplitude encoding}.
\end{principle}

If QN-1 reuses its input qubits and output qubits, the direct connection between QN-1 and QN-2 is feasible. Otherwise, such a connection is infeasible.

The above principle indicates that if QN-1 reuse the qubits for inputs and outputs, we can connect QN-1 and QN-2.
This is because in such a situation (like V-NEU), its output qubits could be interpreted as amplitudes directly and thus becomes the case in Principle \ref{p:path6}. Otherwise, since the output qubits of QN-1 are entangled with the input qubits, the change of the amplitude of one output qubit will affect the amplitude of another output qubit. As a result, we cannot independently compute two neurons in the inference phase, leading to high computation complexity.
This is violate to the second design goal mentioned in Section~\ref{Sec:Method}.

% the output qubits of its previous layer, which could not achieve the second design goal mentioned in Section~\ref{Sec:Method} if the direct connection between QN-1 and QN-2 is applied. 

% Since the probability encoding are initially proposed for the quantum neurons in the first layer to encode the classical input data with the assumption that the data items of input are independent. Based on the third design goal in Section~\ref{Sec:Method}, when it is used by QN-2 to encode the intermediate input qubits, the requirement of independence between the input qubits (i.e., the output qubits of QN-1) still holds. However, since the output qubits of QN-1 are entangled and thus not independent, the direct connection does not satisfy the requirement QN-2 has on its input. 

\begin{principle}\label{p:path1}
\textbf{(Path 8)} 
If output qubits of QN-1 are probability encoding and entangled, and the input of QN-2 is also probability encoding, QN-1 and QN-2 can be connected if the design in QN-2 (1) uses outputs of QN-1 as control end without phase kickback, or (2) operate on the outputs of QN-1 rotates around $X$-axis only (i.e., using RX gate) or the combine of (1) and (2).
% If QN-1 contains quantum gates that entangle its output qubits and its output qubits use \textit{probability encoding}, whether the direct connection between QN-1 and QN-2 is feasible is conditional when the expected encoding method of the input of QN-2 is \textit{probability encoding}.
\end{principle}

Due to the limited space, the detailed proof is omitted here.

As an example, in \net, it connects V-NEU and U-NEU (Path 5, Related to Principle 2), U-NEU and N-NEU (Path 8, related to Principle 5), N-NEU and P-NEU (Path 8). V-NEU and P-NEU will also be connected if $R_2 = 0$. To make the connection between V-NEU and P-NEU feasible, V-NEU has to encode its output as probability (Path 8). Since the implementation of P-NEU shown in Figure~\ref{fig:motivation}(d) does not introduce phase kickback at the end and also not use any rotation gate, the condition of Principle 5 is satisfied. Therefore, all of these connections are feasible, and thus \net could achieve the design goals listed in Section~\ref{Sec:Method}.

% Note although V-NEU could encode its output as probability, we only use amplitude encoding in \net, which follows Principle \ref{p:path6} to connect to U-LRY.

% The connection between U-NEU and N-NEU or P-NEU satisfy the condition proposed by Principle \ref{p:path1}. 

% As shown in Table~\ref{tab:encoding}, U-Neuron applies amplitude encoding to its input and probability encoding to its output. V-Layer uses amplitude encoding to both of its input and output. But it is also feasible to convert the amplitude of the output to probability without quantum-classical communication if we adds appropriate controlled NOT gate (CNOT gate) before sending the output to the next layer. For N-neuron and P-neuron, both of them use probability encoding to process their input and output. 

% \clearpage

\section{Numerical Experiments}\label{Sec:Experiment}
This section reports the related evaluation results of \net on MNIST dataset and its different sub-datasets.
Results demonstrated that the state-of-the-art VQC and QuantumFlow can only obtain 52.77\% and 69.92\% accuracy on MNIST dataset, while \net can achieve over 20\% accuracy gain, reaching 90.62\%.

% We first evaluate the performance of \net by simply using a single real-valued V-Layer. A comparison is also made between \net and baselines (i.e., QF-Net and VQC-based QNN). Then, we conduct sensitivity analysis of the number of V-Layer in \net to explore the impact of this factor on the performance of \net.  

% The experimental results in Section~\ref{Subsec:exp_compare} show that \net could achieve comparable or better performance than the baselines on all of the evaluated datasets. Results from Section~\ref{Subsec:exp_v_layer} demonstrate the potential of \net whose performance could be effectively improved by adding more V-Layers and thus increasing the proportion of the real-valued weights.~\todo{Exploration?}

\noindent\textit{A. Experimental Setting}~\label{Subsec:exp_setting}
% In this section, we will introduce the experimental setting from the aspect of evaluated dataset, the neural architectures of \net and the baselines for comparison.

\begin{table}[t]
\centering
\renewcommand{\arraystretch}{1.3}
\tabcolsep 1pt
\caption{Evaluation of QNNs with Different Neural  Architecture}
\label{tab:MIX-NN Evaluation}
\begin{tabular}{ccccccc}
\hline
\multicolumn{2}{c}{Architecture} & \multicolumn{1}{c}{MNIST-2$^\dag$}  & \multicolumn{1}{c}{MNIST-3$^\dag$} & \multicolumn{1}{c}{MNIST-4$^\ddag$} & \multicolumn{1}{c}{MNIST-5$^\ddag$} & \multicolumn{1}{c}{MNIST$^\S$} \\ \hline
\multicolumn{2}{c}{VQC (V$\times R1$)} & \textbf{97.91\%} & 90.09\% & 93.45\% & 91.35\% & 52.77\% \\ 
\multicolumn{2}{c}{QuantumFlow} & 95.63\% & 91.42\% & 94.26\% & 89.53\% & 69.92\% \\
\hdashline  
\multirow{3}{*}{\net} & V+U & 97.36\% & \textbf{92.77\%} & \textbf{94.41\%} & \textbf{93.85\%} & 88.46\% \\
& V+U+P & 87.45\% & 82.9\% & 92.44\% & 91.56\% & \textbf{90.62\%} \\
& V+P & 91.72\% & 76.93\% & 88.43\% & 85.02\% & 49.57\% \\
\hline
\multicolumn{6}{l}{\footnotesize Input resolutions: $^\dag$ $4\times4$; $^\ddag$ $8\times8$; $^\S$ $16\times16$; }\\
\hline
\end{tabular}
\end{table}

\textbf{Dataset}. We employ different sub-datasets of MNIST dataset to evaluate the performance of different quantum neural architectures for classifying the handwritten digits. A sub-dataset with $X$ classes is denoted as MNIST-$X$, where $X \in \{2, 3, 4, 5, 10\}$. For MNIST-2, we select digits 3 and 6 for the classification, which is denoted as $\{3, 6\}$. Moreover, $\{0,3,6\}, \{0,3,6,9\}, \{0, 1, 3, 6, 9\}$ are the specific chosen sub-datasets for MNIST-3, MNIST-4 and MNIST-5, respectively. 
MNIST-10 represents the original dataset, abbreviated as MNIST. 
% Note that since MNIST dataset has ten classes, MNIST-10 is equivalent to the MNIST dataset itself. 
Before training and inference, images within the dataset should be downsampled to a smaller resolution to reduce the number of input qubits. 
Specifically, the original image is downsampled from the resolution of $28\times 28$ to $4\times 4$, $8\times 8$, $16\times 16$ for different datasets, shown in Table \ref{tab:MIX-NN Evaluation}.

% for MNIST-2 and MNIST-3, while for MNIST-4 and MNIST-5, it is downsampled to . A resolution of $16\times 16$ is applied to images from MNIST.

\textbf{Quantum Neural Architectures}. 
For the intermediate layers in quantum neural architecture, we need to specify the number of neurons in U-Layer and P-Layer.
In the experiments, we set the value to be $4$, $8$, $16$, $16$, and $32$ for MNIST-2, MNIST-3, MNIST4, MNIST-5 and MNIST, respectively.

% We set the number of neurons in U-Layer and P-Layer to be $4$ for MNIST-2 and $8$ for MNIST-3.
% % should be manually decided, in the following experiments,  we set the number to $4$ for MNIST-2 and $8$ for MNIST-3. 
% For MNIST-4 and MNIST-5, the number is set as 16. Besides, $32$ neurons are used in MNIST to get a model with larger capacity.

% Two types of V-Layer are implemented in our experiments. Their detailed designs are shown in Figure~\ref{}.~\todo{Figure of V5 and V10}
% \todo{Two version of VQC} The input and output size of V-Layer could be decided by using the rules we mentioned in Section~\ref{Sec:Method}. 

%  The number of neurons of N-Layer could be obtained in the same way. For U-Layer and P-Layer, the number of neurons of some of them can be got by following the rules in Section~\ref{Sec:Method}. For the remaining of them, we use predefined values. More specifically, 
 
\textbf{Baselines}. The baseline QNNs contain (1) the QNNs from Quantumflow~\cite{jiang2021co} and (2) a variant of the VQC in~\cite{sim2019expressibility}, whose circuit implementation is shown in Figure~\ref{fig:motivation}(b).

% \todo{training hyperparameters?}

\noindent\textit{B. Evaluation of \net}~\label{Subsec:exp_compare}
% \todo{Better title?}

Table~\ref{tab:MIX-NN Evaluation} reports the performance comparison among VQC-based QNN, QuantumFlow, and \net.
In this table, notations V, U, and P represent V-Layer, U-Layer, and P-Layer, respectively. 
Note that we do not show N-Layer in \net explicitly here for simplicity. 
For each architecture, we tried multiple configurations of N-Layer and report the best results. For VQC-based NN, $V\times R1$ represents that the same VQC block is repeated by R1 times to construct the entire VQC, as shown in Figure \ref{fig:method}. We tried the value of $R1$ from 1 to 5, and report the best results in Table~\ref{tab:MIX-NN Evaluation}.

We have several observations from the results, for MNIST-2 (requiring binary classification) a linear decision boundary might be sufficient for classifying, which explains why VQC-based QNN achieves the best result that is $0.55\%$ higher than the best architecture from \net. 
On MNIST-\{3,4,5,10\} that are more complicated than MNIST-2, the best architecture from \net always outperforms VQC-based QNN. 
More specifically, on MNIST, \net can achieve $37.85\%$ higher accuracy over the VQC-based QNNs, which shows that for complicated dataset that is usually not linear separable, increasing the number of linear layers without non-linear function cannot achieve good performance. Therefore, it is important to build a deep QNN with nonlinear function following each linear layer. Besides, the best architecture from \net outperforms QF-Net on all the evaluated datasets consistently. It proves that by adding more real-valued weights to QF-Net, the resulted neural architecture can achieve better performance. Therefore, we can conclude that VQC-based QNN and \net are complementary to each other. By mixing the neurons from them carefully, \net could produce better quantum neural architecture.   

% \begin{table}[]
% \centering
% \caption{Encoding Method of Basic Neurons}
% \label{tab:encoding}
% \begin{tabular}{lcc}
% \hline
% \multicolumn{1}{c}{Neuron Type} & \begin{tabular}[c]{@{}c@{}}Input Encoding\\ Method\end{tabular} & \begin{tabular}[c]{@{}c@{}}Output Encoding \\ Method\end{tabular} \\ \hline
% U-Neuron & Amplitude & Probability \\
% V-Neuron & Amplitude & Amplitude/Probability \\
% P-Neuron & Probability & Probability \\
% N-Neuron & Probability & Probability \\ \hline
% \end{tabular}
% \end{table}

% \todo{do I need to discuss the performance of the three architecture of \net?}

% \begin{table*}[]
% \centering
% \caption{Evaluation of QNN with Different Neural  Architecture}
% \label{tab:MIX-NN Evaluation}
% \begin{tabular}{lccccc}
% \hline
% Architecture & MNIST-2 & MNIST-3 & MNIST-4 & MNIST-5 & MNIST \\ \hline
% V + U (QF-MIXNN) & 97.36\% & \textbf{92.77\%} & \textbf{94.41\%} & \textbf{93.85\%} & 88.46\% \\
% V + U + P (QF-MIXNN) & 87.45\% & 82.9\% & 92.44\% & 91.56\% & \textbf{90.62\%} \\
% V + P (QF-MIXNN) & 91.72\% & 76.93\% & 88.43\% & 85.02\% & 49.57\% \\
% U + P (QF-Net) & 95.63\% & 91.42\% & 94.26\% & 89.53\% & 69.92\% \\
% V (Best R) & \textbf{97.91\%} & 90.09\% & 93.45\% & 93.45\% & 52.77\% \\ \hline
% \end{tabular}
% \end{table*}

\begin{figure}[t]
\centering
\includegraphics[width=1\linewidth]{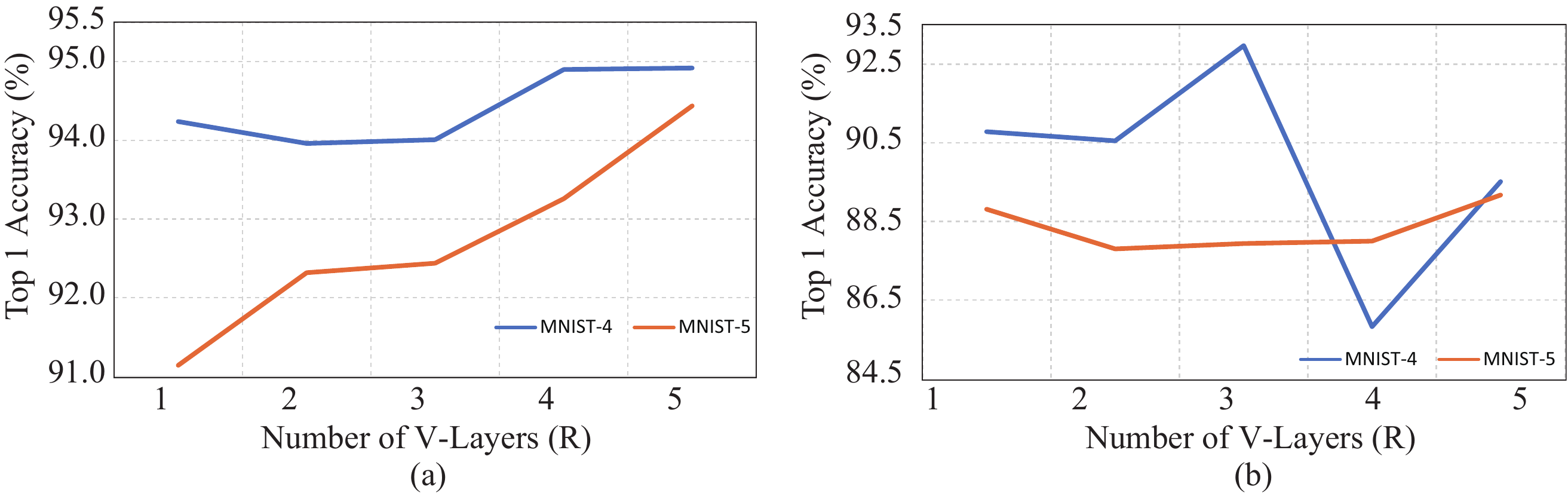}
\vspace{-20pt}
\caption{Model accuracy changes along with the number of repetition blocks in VQC increases: (a) \net-1, $V\times R+U$; (b) \net-2, $V\times R+U+P$ (Best viewed in color)}
\label{fig:senstive}
\end{figure}

\noindent\textit{C. Sensitivity Analysis of Number of V-Layer}~\label{Subsec:exp_v_layer}

We conduct sensitivity analysis on MNIST-4 and MNIST-5 to explore the impact of the number of V-Layer (R) on the performance of \net-1 (i.e., $V\times R+U$) and \net-2 (i.e., $V\times R+U+P$). Figures \ref{fig:senstive}(a)-(b) show the results of \net-1 and \net-2, respectively. As we can see in Figure~\ref{fig:senstive}, when R increases, the accuracy of the model has a trend to improve. 

An interesting observation is obtained for \net-2 on MNIST-4. Along with the increase of R, the accuracy starts to drop when R is greater than 3. Since we use the same training setting for different value of R for fair comparison, this result might indicate that for QNN with more layers like \net-2, we should pay more attention to tune the training parameters if we want to obtain accuracy gain by increasing the number of V-Layers. But even for \net-2 on MNIST-4, we can achieve $2.19\%$ accuracy gain when R increases from 1 to 3. 

In summary, we can conclude that the performance of the neural architectures from \net could be further improved if we increase its total number of real-valued weights by adding more V-Layers.   

% \subsection{Exploration}~\label{Subsec:exp_v_layer}

% \begin{table}[]
% \centering
% \caption{Exploration of Different quantum neural architecture}
% \label{tab:Exploration of Combination}
% \begin{tabular}{lrrrr}
% \hline
% Architecture & \multicolumn{1}{c}{w/o Measurement} & \multicolumn{1}{c}{\# Qubit} & \multicolumn{1}{c}{Accuracy} & \multicolumn{1}{c}{Training Cost} \\ \hline
% V + U &  \checkmark& \textbf{} & \textbf{+++} & \textbf{} \\
% V + U + P &  \checkmark&  & +++ &  \\
% V + P &  \checkmark&  & + &  \\
% U + P &  \checkmark&  & ++ &  \\
% U + V & \textbf{} &  & -- &  \\
% U + V + P & \multicolumn{1}{l}{} & \multicolumn{1}{l}{} & -- & \multicolumn{1}{l}{} \\ \hline
% \end{tabular}
% \end{table}

% \clearpage

% \clearpage
\section{Conclusion}\label{Sec:conclusion}
In this paper, we propose to mix different quantum neuron designs to explore a better quantum neural architecture. A set of general principles to mix different quantum neurons are developed, which can be as a foundation for the future research on quantum neuron mixture.
Based on these design principles, we mix the quantum neurons from VQC-based QNN and QuantumFlow, and identify the \net to take advantage of both design.
As a result, the newly explored \net can 
% \net, a novel quantum neural architecture design by seamlessly mixing the different quantum neurons from VQC-based QNN and Quantumflow. And a set of general principles to mix different quantum neurons are also provided as a foundation for the future research on quantum neuron mixture. The experimental results show that the neural architecture from \net could 
achieve 90.62\% of accuracy on MNIST, which outperforms the neural architectures without our method of mixture by a large margin. 
% Moreover, by increasing the number of V-Layer in \net, the performance of the neural architecture could be improved further.

\section*{Acknowledgment}
Special thanks to Zhirui Hu for her extensive discussion and help for this work.

\clearpage

% Add our own bib section
%{
% \bibliographystyle{IEEEtran}
\bibliographystyle{unsrt}
\bibliography{bibligraphy}

\end{document}